\documentclass[12pt,a4paper]{article}
\usepackage{amsfonts,latexsym}
\usepackage{amsmath,amssymb}
\usepackage{graphicx,color,}

\renewcommand{\thefootnote}{\fnsymbol{footnote}}

\begin{document}

\vspace{12mm}

\begin{center}
{{{\Large {\bf  Analytical approximate solutions for scalarized AdS black holes  }}}}\\[10mm]

{De-Cheng Zou$^{a}$\footnote{e-mail address: dczou@yzu.edu.cn},
Bo Meng$^{a}$\footnote{mb20210111@163.com;},
Ming Zhang$^{b}$\footnote{zhangming@xaau.edu.cn;}, Sheng-Yuan Li$^{a}$\footnote{lishengyuan314159@hotmail.com;},\\
Meng-Yun Lai$^c$\footnote{mengyunlai@jxnu.edu.cn;} and Yun Soo Myung$^d$\footnote{ysmyung@inje.ac.kr;}}\\[8mm]

{${}^a$Center for Gravitation and Cosmology and College of Physical Science and Technology,\\ Yangzhou University, Yangzhou 225009, China\\[0pt]}
{${}^b$Faculty of Science, Xi'an Aeronautical University, Xi'an 710077, China\\[0pt]}
{${}^c$College of Physics and Communication Electronics, Jiangxi Normal University, \\ Nanchang 330022, China\\[0pt] }
{${}^d$Institute of Basic Sciences and Department  of Computer Simulation,\\ Inje University, Gimhae 50834, Korea\\[0pt] }
\end{center}

\vspace{2mm}

\begin{abstract}
The spontaneous scalarization of Schwarzscild-AdS is investigated in the Einstein-scalar-Gauss--Bonnet (ESGB) theory.
Firstly, we construct scalarized  AdS black holes numerically. Secondly, making use of the homotopy analysis method (HAM),
we obtain  analytical approximate solutions for scalarized AdS black holes in the ESGB theory.
It is found that scalarized  AdS black holes constructed numerically are consistent with  analytical approximate solutions in  the whole space.
\end{abstract}
\vspace{5mm}

\newpage
\renewcommand{\thefootnote}{\arabic{footnote}}
\setcounter{footnote}{0}


\section{Introduction}

In general relativity (GR), the ``no-hair theorem'' has always been a hot topic. It allows that a GR
black hole can be described by  three observables of mass $M$, electric charge $Q$,
and rotation parameter $a=J/M$ \cite{Carter1971zc,Ruffini1971bza}, and rules out a black hole
coupled to a scalar field in asymptotically flat spacetimes, on account of the
divergence of scalar field on the horizon \cite{Bekenstein:1974sf,Bekenstein:1975ts,Bronnikov:1978mx}.
In the 1990s, Damour and Esposito-Farese~\cite{Damour:1993hw,Damour:1996ke} have first found a new
mechanism of spontaneous scalarization
in scalar-tensor theory in neutron stars.
This phenomenon has received a lot of attention lately. Considering a scalar field function $f(\phi)$
coupling to the Gauss--Bonnet curvature term $R^2_{\rm GB}$ such as $f(\phi) R^2_{\rm GB}$~\cite{Doneva:2017bvd,Silva:2017uqg,Antoniou:2017acq,Myung:2018iyq},
scalarized black hole solutions were found in ESGB theory, where the coupling term causes instability near
the event horizon of a Schwarzschild black hole and induces scalarized
black holes. Then, the so-called ``no-hair theorem'' of GR \cite{Bekenstein:1995un} can be avoided in ESGB theory.
It is worth pointing out that, in ESGB theory, there is no a priori guidance for determining the coupling function $f(\phi)$.
The coupling function $f(\phi)$ has a decisive influence on the properties of the scalarized black holes.
For instance, Ref.~\cite{Doneva:2017bvd} adopted the exponential coupling $f(\phi)\sim\exp(\beta\phi^2)$,
while Ref.~\cite{Silva:2017uqg} focused on the quadratic coupling $f(\phi)\sim \beta\phi^2$ instead.
These theories possess black holes with scalar hair, whose
properties have been investigated in great detail~\cite{Minamitsuji:2018xde,Doneva:2018rou,Macedo:2019sem,Blazquez-Salcedo:2021npn,Doneva:2021tvn}.
In addition, Ref.~\cite{Blazquez-Salcedo:2018jnn} has noticed that, under radial perturbations,
the scalarized black holes are unstable for a quadratic coupling, whereas it is stable for
an exponential form in the ESGB theory. Motivated by current and future
gravitational wave observations from black hole mergers, the axial~\cite{Blazquez-Salcedo:2020rhf}
and polar~\cite{Blazquez-Salcedo:2020caw} perturbations of scalarized black holes have
been investigated  to obtain the quasinormal modes (QNMs) in the ESGB theory since
QNMs could describe the ringdown after merging.

It is well-known that the anti-de Sitter/conformal field theory (AdS/CFT)
correspondence provides a powerful framework for studying quantum mechanical aspects of black hoes~\cite{Maldacena:1997re,Gubser:1998bc}.
In some scenarios, holographic duality has allowed us to bring CFT knowledge to bear on black hole physics in asymptotically AdS space-time.
Moreover, a scalar field in an asymptotically AdS space-time can cause an asymptotic instability only if its
mass-squared $\mu^2_{\rm eff}$ is less than the BF bound $\mu^2_{\rm BF}$~\cite{Breitenlohner:1982jf}.
Then, the SAdS black hole may evolve to a scalarized AdS black hole through tachyonic instability,
and the ``no-hair theorem'' can usually be circumvented.
Bakopoulos et al. \cite{Bakopoulos:2018nui} have firstly discussed the emergence of novel,
regular black hole solutions in ESGB theory. Recently, the scalarization of AdS black holes with applications to holographic
phase transitions was studied in Einstein-scalar-Ricci-Gauss--Bonnet gravity~\cite{Brihaye:2019dck}.
In addition,  Guo et al. have discussed the holographic realization of scalarization in the ESGB
gravity with a negative cosmological constant~\cite{Guo:2020sdu}, and a horizon curvature
has an effect on the scalarization~\cite{Guo:2020zqm}.

Nevertheless, the numerical black hole solutions were obtained at fixed values of parameters. From these numerical solutions,
it is usually hard to give a clear picture for dependence of the metric on physical parameters of the system.
Moreover, these numerical solutions are displayed by some curves in figures, instead of expressions in explicit form.
It causes these solutions of scalarized black hole to usually need to be re-calculated by colleagues in some relevant research work.
Fortunately, the general methods for parametrization of the black hole space-times (continued fractions method (CFM)~\cite{Rezzolla:2014mua}
and homotopy analysis method (HAM)~\cite{Liao:1992mua,Liao:2003mua}) were developed. The CFM
 has recently been applied with success in a variety of contexts \cite{Konoplya:2016jvv}-\cite{Sajadi:2022ybs}.
We stress here that the HAM is also a very powerful method for obtaining analytical approximate solutions to various nonlinear
differential equations (including systems of nonlinear equations and arising in many different areas of science
and engineering~\cite{Xu:2010}-\cite{Zhang:2019}. Despite its popularity in many areas of science and engineering over the years,
the application of the HAM has been very limited in the fields of general relativity and gravitation.
Recently, this HAM has been adopted to derive analytic approximate solutions of field equations
in Einstein--Weyl gravity~\cite{Sultana:2019lhf,Sultana:2021cvq} as well as  analytic expression
of Regge--Wheeler equations under the metric perturbations on Schwarzschild space-time~\cite{Cho:2020tzx}.
In this work, firstly, we construct scalarized AdS black holes numerically.  Secondly, making use of the HAM,
we wish to obtain analytical approximate solutions for scalarized AdS black holes in the ESGB theory.

The plan of our work is as follows. In Section~\ref{sec2}, we investigate the tachyonic instability of
Schwarzschild AdS (SAdS) black holes under the linearized scalar perturbation in the ESGB theory.
Then, we construct numerical solutions of scalarized AdS black holes in Section~\ref{sec3}.
Section~\ref{sec4} is devoted to deriving analytical approximation solutions by introducing the HAM,
where two solutions are accurate in the whole space outside the event horizon.
Finally, we end the paper with a discussion and conclusions in Section~\ref{sec5}.

\section{Instability of SAdS black hole}  \label{sec2}

The action for ESGB theory with a negative cosmological constant $\Lambda$ is given by
\begin{equation}
S_{\rm ESGBC}=\frac{1}{16 \pi}\int d^4 x\sqrt{-g}\left( R-2\Lambda-2\partial_\mu \phi \partial^\mu \phi
+ \frac{\lambda^2\phi^2}{2} R^2_{\rm GB}\right),\label{Action1}
\end{equation}
where $\lambda$ is the scalar coupling constant, $R$ the Ricci scalar, $\phi$ a scalar field, and ${\cal R}^2_{\rm GB}$ the Gauss--Bonnet term
\begin{equation}
{\cal R}^2_{\rm GB}=R^2-4R_{\mu\nu}R^{\mu\nu}+R_{\mu\nu\rho\sigma}R^{\mu\nu\rho\sigma}\label{GBterm}
\end{equation}
with Ricci tensor $R_{\mu\nu}$ and Riemann tensor $R_{\mu\nu\rho\sigma}$.

Varying the action (\ref{Action1}) with scalar $\phi$ and metric $g_{\mu\nu}$, one obtains the scalar field equation
\begin{eqnarray}
\square \phi +\frac{\lambda^2}{4} R^2_{\rm GB} \phi=0 \label{s-equa}
\end{eqnarray}
and Einstein equation
\begin{eqnarray}
 G_{\mu\nu}=\Lambda g_{\mu\nu}+2\partial _\mu \phi\partial _\nu \phi -(\partial \phi)^2g_{\mu\nu}
 -2\lambda^2 \nabla^\rho \nabla^\sigma (\phi^2)P_{\mu\rho\nu\sigma}, \label{equa1}
\end{eqnarray}
where $G_{\mu\nu}=R_{\mu\nu}-(R/2)g_{\mu\nu}$ is the Einstein tensor, and  $P_{\mu\rho\nu\sigma}$ is given by
\begin{eqnarray}
P_{\mu\rho\nu\sigma}&=&R_{\mu\rho\nu\sigma}+g_{\mu\sigma}R_{\nu\rho}-g_{\mu\nu}R_{\rho\sigma}+g_{\nu\rho}R_{\mu\sigma}-g_{\rho\sigma}R_{\mu\nu}
+\frac{R}{2}(g_{\mu\nu}g_{\rho\sigma}-g_{\mu\sigma}g_{\nu\rho}). \label{equa2}
\end{eqnarray}
Topological black holes are found without scalar hair as
\begin{equation}
ds^2_{\rm SAdS}=-f_k(r)dt^2+\frac{1}{f_k(r)}dr^2+r^2\left(d\theta^2+\sin^2\theta d\varphi^2\right)\label{metric}
\end{equation}
with
\begin{equation}
f_k(r)=k-\frac{2M}{r}-\frac{\Lambda r^2}{3},
\end{equation}
where $\Lambda=-3/L^2$ with $L$ the curvature radius of AdS space-time.  The cases of $k=0,-1$ were discussed in~\cite{Guo:2020zqm}.
Here, afterwards, we choose the $k=1$ case of
\begin{equation}
f(r)=1-\frac{2M}{r}-\frac{\Lambda r^2}{3} \label{SAdS-m}
\end{equation}
which corresponds to the SAdS black hole.
From $f(r_h)=0$, the outer horizon radius $r_h$ of SAdS black hole is obtained as
\begin{eqnarray}
{r_h}=-\frac{1}{{\left({3M{\Lambda ^2}+\sqrt{9{M^2}\Lambda^4-\Lambda ^3} } \right)}^{1/3}}
-\frac{{\left({3M{\Lambda^2}+\sqrt{9{M^2}\Lambda^4-\Lambda ^3} } \right)}^{1/3}}{\Lambda},\label{radius}
\end{eqnarray}
where the horizon radius $r_h>0$ is always satisfied on account of a positive mass $M>0$ of a black hole
and a negative cosmological constant $\Lambda<0$. Moreover, the mass of SAdS black hole is determined as
\begin{eqnarray}
M=\frac{1}{6}r_h\left(3-\Lambda r_h^2\right).\label{M}
\end{eqnarray}

Now, we discuss the dynamical stability, Breitenlohner--Freedman (BF) bound, and tachyonic instability of SAdS black hole in the ESGB theory.
For this purpose, we need to consider two linearized equations which describe the propagation of metric perturbation $h_{\mu\nu}$ and scalar perturbation
$\delta \phi$
\begin{eqnarray}
 && \delta R_{\mu\nu}(h) =\frac{\bar{g}_{\mu\nu}}{2}\delta R+ \Lambda h_{\mu\nu}, \label{l-eq1} \\
 && \bar{\square}\delta \phi -\mu^2_{\rm eff} \delta \phi= 0, \label{l-eq2}
\end{eqnarray}
which are obtained by linearizing Equations~(\ref{s-equa}) and (\ref{equa1}). As was pointed out in Refs.~\cite{Cardoso:2001bb,Ishibashi:2003ap,Moon:2011sz},
it is clear that the SAdS black hole is dynamically stable when making use of the Regge--Wheeler prescription under metric perturbation.
In an asymptotically AdS space-time, a scalar field can cause an asymptotic instability only if its mass-squared $\mu^2_{\rm eff}$
is less than the BF bound $\mu^2_{\rm BF}=-\frac{9}{4L^2}\equiv\frac{3\Lambda}{4}$~\cite{Breitenlohner:1982jf}.
One always finds $\mu^2_{\rm eff}>\mu^2_{\rm BF}$ for large enough $r$ and
thus the SAdS black hole is stable asymptotically against the formation of the scalar field.
However, if $\mu^2_{\rm eff}<\mu^2_{\rm BF}$ in the intermediate region, the SAdS black hole may evolve to a scalarized
AdS black hole through tachyonic instability. In our case, the effective mass $\mu^2_{\rm eff}$ is fixed as
\begin{eqnarray}
\mu^2_{\rm eff}=-\frac{\lambda^2}{4}\bar{R}^2_{\rm GB}=-\frac{2\lambda^2\Lambda^2}{3}-\frac{12\lambda^2M^2}{r^6}
\end{eqnarray}
and the condition for asymptotic instability  is obtained as
\begin{eqnarray}
&&\mu^2_{\rm eff}<\mu^2_{\rm BF}:\quad -\frac{2\lambda^2 \Lambda^2}{3}<\frac{3\Lambda}{4}\rightarrow \Lambda>-\frac{9}{8}\lambda^2.
\end{eqnarray}

Now, we are in a position to perform the numerical analysis for the tachyonic instability of SAdS black hole in the ESGB theory.
Taking into account the separation of variables,
\begin{eqnarray}
\delta \phi(t,r,\theta,\varphi)=\frac{\psi(r)}{r}Y_{lm}(\theta,\varphi)e^{-i\omega t},
\end{eqnarray}
and introducing a tortoise coordinate $dr_*=dr/(1-2M/r-\Lambda r^2/3)$,
the radial part of Equation (\ref{l-eq2}) is given by
\begin{eqnarray}\label{perteq}
\frac{d^2\psi}{dr_*^2}+\Big[\omega^2-V_{\rm eff}(r)\Big]\psi(r)=0,
\end{eqnarray}
where the effective potential $V_{\rm eff}(r)$ takes the form
\begin{eqnarray} \label{pot-c}
V_{\rm eff}(r)=\Big(1-\frac{2M}{r}-\frac{\Lambda r^2}{3}\Big)\Big[\frac{2M}{r^3}+\frac{l(l+1)}{r^2}
-\frac{2\Lambda}{3}\Big(1+\lambda^2\Lambda\Big)-\frac{12\lambda^2M^2}{r^6}\Big].
\end{eqnarray}
In the next sections, we only consider the case of $l=0$.

\begin{figure*}[t!]
\centering
\includegraphics{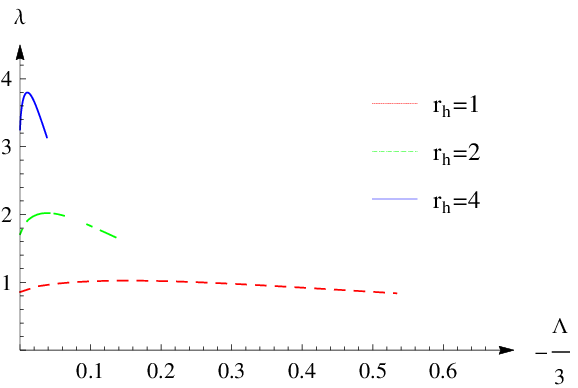}
 \hfill%
\includegraphics{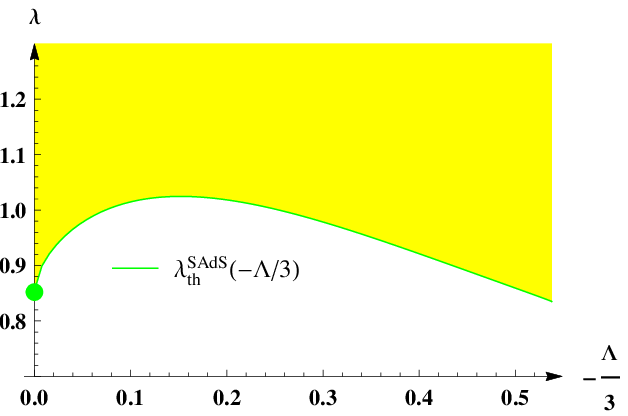}
\caption{(Left) The existence curve for scalarized AdS black holes (threshold curve $\lambda_{\rm th}^{\rm SAdS}(-\Lambda/3)$ of tachyonic instability) in the $(-\Lambda/3,\lambda)$ plane for three different horizon radii $r_h=1,2,4$; (Right) the unstable region is plotted for the horizon radius $r_h=1$ of SAdS black holes.\label{fig1}}
\end{figure*}

To determine the threshold of tachyonic instability, one has
to solve the second-order differential equation numerically
\begin{eqnarray}
\frac{d^2\psi}{dr_*^2}-\Big[\Omega^2+V_{\rm eff}(r)\Big]\psi(r)=0,\label{Omegaeq}
\end{eqnarray}
which allows an exponentially growing mode of $e^{\Omega t}$ $(\omega=i\Omega, \Omega>0)$ as an unstable mode.
Considering $\Omega=0$, we may solve the static linearized equation
\begin{eqnarray}
\frac{d^2\psi}{dr_*^2}-V_{\rm eff}(r)\psi(r)=0,\label{staticeq}
\end{eqnarray}
to find out the threshold unstable mode propagating around the fixed SAdS black hole background. To impose the boundary conditions, we first consider the near-horizon expansion,
which is used to set data outside the horizon for a numerical integration to near infinity
\begin{eqnarray}
\psi(r)=\sum_{i\geq0}\psi_{i}(r-r_h)^i.\label{horizon1}
\end{eqnarray}
In the asymptotic far region, Equation~\eqref{staticeq} becomes approximately
\begin{eqnarray}
\psi''(r)+\frac{2}{r}\psi'(r)-\frac{2+2\lambda^2\Lambda}{r^2}\psi(r)\approx0.\label{infinity1}
\end{eqnarray}
Then, we can obtain the boundary condition of  $\psi(r)\sim r^{-\frac{1}{2}\pm\frac{1}{2}\sqrt{9+8\lambda^2\Lambda}}$ at large $r$.
Therefore, the numerical solution
to Equation~\eqref{staticeq} can be performed by using the shooting method in the region between the black hole horizon and infinity,
seeking for a value of the eigenvalue $\lambda$. These solutions are labelled by an integer $n\in\mathbb{N}_0$: $n=0$ is the fundamental mode,
whereas $n>1$ are excited states (overtones).
We focus on the fundamental mode since the fundamental solutions is usually stable.
Varying $-\Lambda/3$, a set of bifurcation points constitutes the existence curve (threshold curve for tachyonic instability). Figure~\ref{fig1}a includes
three threshold curves of $r_h=1,2,4$. If one chooses $r_h=1$, the unstable region is the upper of threshold curve while the stable
region is the lower of threshold curve; see Figure~\ref{fig1}b.
In case of $-\Lambda/3\rightarrow0$, the value of coupling parameter $\lambda$ matches the threshold value ($\lambda_{\rm th}^{\rm S}=0.852,1.704,3.408$) for the
fundamental mode of the Schwarzschild black hole in~\cite{Silva:2017uqg,Myung:2018iyq}.
This result naturally leads to the fact that the SAdS black hole is unstable in the upper region and thus there exist scalarized AdS black holes in the ESGB theory.

\vspace{-12pt}

\section{Numerical Solutions for Scalarized AdS Black Holes} \label{sec3}

We consider static and spherically symmetric space-times as well as static and spherically symmetric scalar field configuration. The space-time metric and scalar are chosen to be
\begin{equation}
ds^2=-A(r)dt^2+\frac{1}{B(r)}dr^2+r^2\left(d\theta^2+\sin^2\theta d\varphi^2\right),\quad \phi=\phi(r).\label{ms}
\end{equation}
Now we try to find the numerical solutions for scalarized AdS black hole in the ESGB theory. For this purpose,
we first introduce a coordinate transformation of $z=\frac{r_h}{r}$ so that the metric functions can be derived in the compact region of $0 \leq z\leq 1$, and $A(r)$ and $B(r)$ become $A=A(z)$ and $B=B(z)$.
{Therefore, $z = 0$ always corresponds to infinity ($r\rightarrow\infty$),
and $z=1$ naturally corresponds to the event horizon $r=r_h$ of the black hole. To utilize the threshold values for an unstable region in Figure~\ref{fig1}b,
we will choose $r_h=1$ for the horizon radius of the black hole in the following numerical calculation.}

On the other hand,
the metric functions $A(r)$ and $B(r)$ in Equation \eqref{ms} approach $r^2$ as $r\rightarrow\infty$. In other words,
the new metric functions $A(z)$ and $B(z)$ with $1/z^2$ are divergent at $z=0$. Then, we can further define new metric functions
\begin{eqnarray}
A_z(z)\rightarrow  z^2 A(z),\quad B_{z}(z)\rightarrow z^2B(z) \label{metricz}
\end{eqnarray}
so that the new functions $A_z(z)$ and $B_{z}(z)$ are always regular in the whole region of $0 \leq z\leq 1$.
Fortunately, the scalar field $\phi(z)$ is always regular in the whole region under the coordinate transformation $z=\frac{r_h}{r}$. Then, we set
\begin{eqnarray}
\phi_{z}(z)\rightarrow \phi(z). \label{phiz}
\end{eqnarray}

Substituting the new metric functions Equation~\eqref{metricz} and scalar field  Equation~\eqref{phiz} into Equations \eqref{equa1} and \eqref{equa2}, we have
\begin{eqnarray}
eq_1&=&zB_{z}A_{z}'\left[-r_{h}^2+2z(z^2-r_{h}^2\Lambda-3B_{z})\phi_{z}\phi_{z}'\right]+A_{z}\left[r_{h}^2(-z^2+r_{h}^2\Lambda-2z^2\Lambda\phi_{z}B_{z}'\phi_{z}')\right.\nonumber\\
&&\left.-B_{z}\left(r_{h}^2(-3+z^2(1+4\Lambda)\phi_{z}'^2)+4z\phi_{z}((z^2-2r_{h}^2\Lambda)\phi_{z}'+r_{h}^2z\Lambda\phi_{z}'')\right)\right.\nonumber\\
&&\left.+12zB_{z}^2\phi_{z}\phi_{z}'\right]=0,\label{eq1z}
\end{eqnarray}
\begin{eqnarray}
eq_2&=&2r_{h}^2z^2\Lambda B_{z}\phi_{z}A_{z}'\phi_{z}'-A_{z}\Big\{-r_{h}^2z^2+r_{h}^4\Lambda-r_{h}^2zB_{z}'+2z^4\phi_{z}B_{z}'\phi_{z}'-2r_{h}^2z^2\Lambda\phi_{z}B_{z}'\phi_{z}'\nonumber\\
&&-4zB_{z}^2\Big[z\phi_{z}'^2+\phi_{z}(-\phi_{z}'+z\phi_{z}'')\Big]+B_{z}\Big[3r_{h}^2+z^2(r_{h}^2+4z^2-4r_{h}^2\Lambda)\phi_{z}'^2\notag\\
&&+2z\phi_{z}((2z^2+4r_{h}^2\Lambda-3zB_{z}'\phi_{z}+2z(z^2-r_{h}^2\Lambda)\phi_{z}'')\Big]\Big\}=0,\label{eq2z}
\end{eqnarray}
\begin{eqnarray}
eq_3&=&z^2 \lambda^2 (z^2-B_z) B_z\phi_z A_z'^2+z A_z\Big[-z^3\lambda^2\phi_z A'_z B'_z+2\lambda^2 B_z^2\phi_z(-3A'_z+z A''_z)+z B_z r_h^2 A'_z\phi'_z \nonumber\\
&&+\lambda^2z B_z\phi_z \left(A'_z(2z+3B'_z)-2 z^2 A''_z\right)\Big]+A_z^2\Big[12 \lambda^2 B_z^2 \phi_z+z^2 B'_z (2 z \lambda^2 \phi_z+r_h^2\phi'_z)\nonumber\\
&&-2z B_z\left(\lambda^2\phi_z(2z+3 B'_z)+r_h^2 (2\phi'_z-z \phi''_z)\right)\Big]=0,\label{eq3z}
\end{eqnarray}
where primes denote derivatives with respect to $z$.

In order to obtain the asymptotic form of scalarized AdS black holes, we solve three Equations \eqref{eq1z}--\eqref{eq3z} numerically via a shooting method.
Spherically symmetric black holes have an event horizon $(z=1)$, where the metric functions $A_z$ and $B_z$ vanish, and the scalar field $\phi_z$ tends to a constant:
\begin{eqnarray}\label{expansion}
&&A_{z}(z\approx 1)=A_{1}(1-z)+A_{2}(1-z)^2+\cdots, \\
&&B_{z}(z\approx 1)=B_{1}(1-z)+B_{2}(1-z)^2+\cdots,\\
&&\phi_{z}(z\approx 1)=\phi_{0}+\phi_{1}(1-z)+\cdots,
\end{eqnarray}
where $\phi_{0}$ denotes the scalar field at the horizon. It is worth pointing out that the regularity of a scalar field, and its first
and second derivatives on the horizon give an additional  condition
\begin{eqnarray}
r_h^6-8r_h^4\lambda^4\Big[3+2r_h^2\Lambda\left(r_h^2\Lambda-2\right)\Big]\phi_{0}^2
-48r_h^2\lambda^8\Lambda\left(r_h^2\Lambda-2\right)
\phi_{0}^4>0,\label{horzcond2}
\end{eqnarray}
which reduces to that for the Schwarzschild black hole in the limit of $\Lambda\rightarrow0$ \cite{Doneva:2017bvd}.

\begin{figure*}[t!]
\includegraphics{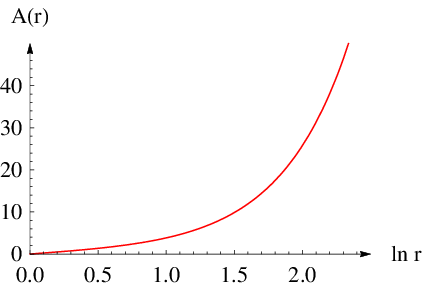}
 \hfill%
\includegraphics{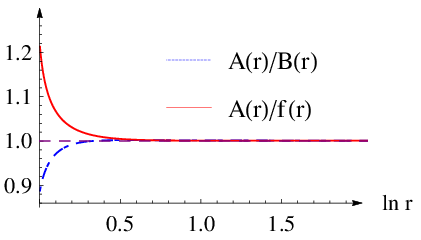}
 \hfill%
\includegraphics{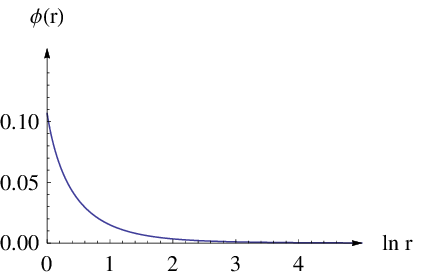}
\caption{The scalarized AdS black hole with $\lambda=0.892$ and $-\Lambda/3=0.457$ belonging to the
fundamental branch of $\lambda>\lambda_b=0.886$ (bifurcation point).
Here, $f(r)$ represents the metric function (\ref{SAdS-m}) for the SAdS black hole.}\label{fig2}
\end{figure*}

On the other hand, the metric functions and scalar field at the infinity $(z\rightarrow 0)$ should satisfy the following boundary conditions:
\begin{eqnarray}
A_z=B_z=-\frac{\Lambda r_h^2}{3}, \quad \phi_z= 0, \quad  {\rm when} \quad z\rightarrow 0 \quad (r\to\infty).
\end{eqnarray}\label{infcond}
We fix $r_h=1$ for the horizon radius of the black hole during the numerical calculation. By tunneling the coupling parameter $\lambda$ and choosing
different values of cosmological constant $\Lambda$, we can obtain a nontrivial solution of scalarized AdS black holes in the ESGB gravity.
The numerical solution for fundamental branch is obtained by taking $\lambda=0.892$ and $-\Lambda/3=0.457$ (greater than 0.886 of bifurcation
point) (see Figure~\ref{fig2}). We plot all figures in terms of $\ln r$ and thus the horizon is always located at $\ln r_h=0$.
Here, $f(r)$ represents the metric function for the SAdS black hole with $\phi_{\rm SAdS}(r)=0$. Notice that the metric
functions $A(r)$ and $B(r)$ display different behaviors in comparison to those for the SAdS black hole and these approach
the SAdS metric function $f(r)$ as $\ln r$ increases. Moreover, a scalar field $\phi(r)$
is a decreasing function with starting with 0.107, and its asymptotic value is zero.

\section{Analytical approximate solutions} \label{sec4}

In general, it is a difficult task to find exact solutions of nonlinear differential equations. In Refs.~\cite{Liao:2003mua,Liao:20041},
the  HAM was developed to obtain analytical approximate solutions to nonlinear differential equations.
Here, we wish to derive analytical approximate solutions for metric functions $A_z(z)$, $B_z(z)$ and a scalar
field $\phi_z(z)$ by solving nonlinear  \mbox{Equations \eqref{eq1z}--\eqref{eq3z}} by using the HAM.
If we succeed to find them, it will confirm the numerical solutions in the previous section.

We assume the nonlinear operators $N_i$, which are suitable for a system of $n$-nonlinear differential equations
\begin{eqnarray}
  N_i[y_i(t)]=0, \qquad\quad i=1,2,...,n, \label{neq}
\end{eqnarray}
with unknown function $y_i(t)$ and a variable $t$. Then,
the zero-order deformation equation can be written as~\cite{Liao:2003mua,Liao:20041}
\begin{eqnarray}
(1-q)L[\phi_i(t;q)-y_{i0}(t)]=q h_i H_i(t)N_i[\phi_i(t;q)] \label{hm}
\end{eqnarray}
where $L$ is an auxiliary linear operator with the property $L[0]=0$, $q\in[0,1]$
is an embedding parameter in topology (called the homotopy parameter),
$\phi_i(t;q)$ are the solutions of Equation \eqref{hm} for $q\in[0,1]$, $y_{i0}(t)$ is the initial guesses,
and $h_i\neq 0$ is the so-called ``convergence-control parameters''.
Considering the property $L[0]=0$, the solutions $\phi_i(t;q)$ of Equation \eqref{hm} vary continuously from
the initial guess $y_{i0}(t)$ to the actual solution $y_i(t)$ of Equation \eqref{neq} when the parameter $q$
increases from $0$ to $1$. Here, we set the auxiliary functions $H_i(t)=1$ without any restrictions.

On the other hand, we can also expand $\phi_i(t;q)$ as the Maclaurin series with respect to $q$
\begin{eqnarray}
  \phi_i(t;q)=y_{i0}(t)+\sum_{m=1}^{\infty}y_{im}(t)q^m, \qquad  y_{im}(t)=\frac{1}{m!}\frac{\partial^m \phi_i(t;q)}{\partial q^m}. \label{macs}
\end{eqnarray}
The proper choice of the initial approximation $y_{i0}(t)$, linear operator $L$, and convergence control parameter $h_i$
will make the series expansion \eqref{macs} convergency at $q=1$. Therefore, we obtain
\begin{eqnarray}
  y_i(t)=\phi_i(t;1)=y_{i0}(t)+\sum_{m=1}^{\infty}y_{im}(t).\label{yit}
\end{eqnarray}
Here the function $y_{im}(t)$ could be obtained by solving the $m^{th}$ order deformation equation. Differentiating Equation \eqref{hm} $m$
times with respect to the parameter $q$, setting $q=0$, and dividing by $m!$, we find the $m$th order deformation equation
\begin{eqnarray}
  L[y_{im}(t)-\chi_my_{im-1}(t)]=h_i R_{im}(y_{im-1}), \label{leq}
\end{eqnarray}
where
\begin{eqnarray}
  R_{im}(y_{im-1})=\frac{1}{(m-1)!}\frac{\partial^{m-1} N_i[\phi_i(t;q)]}{\partial q^{m-1}}\mid_{q=0},
\end{eqnarray}
and
\begin{eqnarray}
  \chi_m=\left\{
  \begin{aligned}
    0 & : m\leq 1 \\
    1 & : m>1.
  \end{aligned}
  \right.
\end{eqnarray}
We define the partial sum $y_i^M(t)$ by
\begin{eqnarray}
  y_i^M(t)=y_{i0}(t)+\sum_{m=1}^{M}y_{im}(t),
\end{eqnarray}
where $y_i^M(t)$ are the $M^{th}$ order approximate solutions of the original Equation \eqref{neq}.

In order to solve Equations \eqref{eq1z}--\eqref{eq3z} by means of the HAM, we choose the initial approximations
\begin{eqnarray}
&&A_{z0}(z)=B_{z0}(z)=\left(z^2-\frac{r_h^2\Lambda}{3}+\frac{z^3}{3}(-3+r_h^2\Lambda)\right)(1-\alpha z),\label{init1}\\
&&\phi_{z0}(z)=\frac{107}{1000}\left(\frac{72}{100}z^3+\frac{28}{100}z\right)\label{init2}
\end{eqnarray}
with an undetermined constant $\alpha$ and corresponding auxiliary linear operators~\cite{Gorder:2009}
\begin{eqnarray}
L[\phi_z]=\frac{d^2\phi_z}{dz^2}, \quad  L[B_z]=\frac{dB_z}{dz}, \quad L[A_z]=\frac{d^2A_z}{dz^2}.
\end{eqnarray}
One can find that the chosen approximations satisfy the initial and boundary conditions, since $A_{z0}$ and $B_{z0}$ vanish at the event horizon $(z=1)$, and they reduce to $-\frac{r_h^2\Lambda}{3}$ as $z\rightarrow0$. Moreover, the scalar field $\phi_{z0}(z)$ disappears at infinity and equals $0.107$ near the horizon $(z=1)$ .

Then, we use the HAM to secure analytical approximations for Equations \eqref{eq1z}--\eqref{eq3z} by using the boundary conditions
\begin{eqnarray}
A_z(0)=-\frac{r_h^2\Lambda}{3},~~A_z(1)=B_z(1)=0,~~\phi_z(0)=0.107,~~\phi_z(1)=0,\label{eqbc}
\end{eqnarray}
where we reserve one boundary condition $B_z(0)=-\frac{r_h^2\Lambda}{3}$ for later computations. The $M$th order approximations of $A_z, B_z, \phi_z$ are  written as
\begin{eqnarray}
  &&A_z(\alpha, h_i, z)\approx A_{z0}(\alpha, z)+\sum_{k=1}^{M}A_{zk}(\alpha, h_i, z), \label{eqapp1a}\\
  &&B_z(\alpha, h_i, z)\approx B_{z0}(\alpha, z)+\sum_{k=1}^{M}B_{zk}(\alpha, h_i, z), \label{eqapp1b}\\
  &&\phi_z(\alpha, h_i, z)\approx \phi_{z0}(z)+\sum_{k=1}^{M}\phi_{zk}(\alpha, h_i, z), \label{eqapp1p}
\end{eqnarray}
which  include  the unknown parameter $\alpha$ and the convergence-control parameter $h_i$.

Considering the boundary condition $B_z(0)=-\frac{r_h^2\Lambda}{3}$ with $M$th order approximate expression \eqref{eqapp1a}, one obtains
\begin{eqnarray}
  \Gamma_M(\alpha, h_i)\equiv B_{z0}(\alpha, 0)+\sum_{k=1}^{M}B_{zk}(\alpha, h_i, 0)+\frac{r_h^2\Lambda}{3}=0 \label{eqgamma},
\end{eqnarray}
where $\Gamma_M$ represents an expanded form of the constrained boundary condition.
As long as $h_i$ is given, a solution to Equation \eqref{eqgamma} is easily obtained. We use the technique developed by Xu et al.~\cite{Xu:2010}
to find out the optimal values of $h_i$. In principle, the technique seeks for minimizing   averaged square residual
error of Equations \eqref{eq1z}--\eqref{eq3z} at the $m$th order
\begin{eqnarray}
E_m(\alpha, h_i)&=&E^{N_1}_m+E^{N_2}_m+E^{N_3}_m \nonumber \\
&=&\frac{1}{S+1}\sum_{k=0}^{S}\bigg[\bigg(N_1[\sum_{n=0}^{m}A_{zn}(z_k),\sum_{n=0}^{m}B_{zn}(z_k),\sum_{n=0}^{m}\phi_{zn}(z_k)]\bigg)^2 \nonumber \\ &&+\bigg(N_2[\sum_{n=0}^{m}A_{zn}(z_k),\sum_{n=0}^{m}B_{zn}(z_k),\sum_{n=0}^{m}\phi_{zn}(z_k)]\bigg)^2 \nonumber\\
&&+\bigg(N_3[\sum_{n=0}^{m}A_{zn}(z_k),\sum_{n=0}^{m}B_{zn}(z_k),\sum_{n=0}^{m}\phi_{zn}(z_k)]\bigg)^2 \bigg]\label{Em}
\end{eqnarray}
with
\begin{eqnarray}
  z_k=k\Delta z=\frac{k}{S}, ~~k=0,1,2,\cdots,S.
\end{eqnarray}
We choose $S=40$ used with the purpose of optimization for each function. For our problem, the residual error depends  on both $\alpha$ and $h_i$. In fact, both $E_m(\alpha, h_i)$ and $\Gamma_M(\alpha, h_i)$ contain undetermined parameters: $\alpha$ and $h_i$. Therefore, the optimal convergence-control parameters $h_i$ can be determined from the minimum of $E_m(\alpha, h_i)$, and it is subjected additionally to the algebraic Equation \eqref{eqgamma} which needs to secure the constant $\alpha$. Mathematically, this doubly coupled optimization problem implies
\begin{eqnarray}
  (\alpha^*, h_i^*)=\text{min}\{E_m(\alpha, h_i),\Gamma_M(\alpha, h_i)=0\}.
\end{eqnarray}

\begin{figure*}[t!]
\centering
\includegraphics{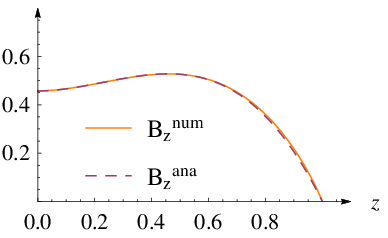}
 \hfill%
\includegraphics{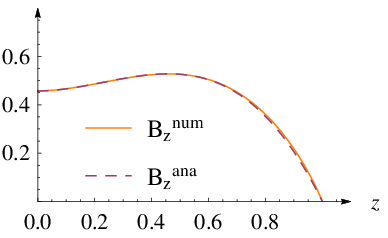}
 \hfill%
\includegraphics{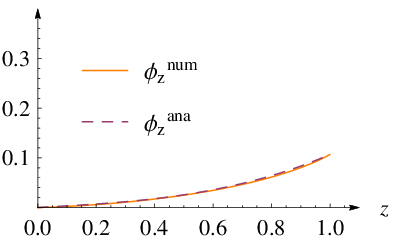}
\caption{Comparison figures of metric functions $A_z$, $B_z$ and scalar field $\phi_z$ in the numerical (solid curve)
and analytical approximate (dashed curve)  solutions. Here, we choose horizon radius parameter $r_h=1$,
$\lambda=0.892$ and $-\Lambda/3=0.457$. }\label{fig33}
\end{figure*}

Considering the 2nd order $(M=2)$ approximation, we obtain
$h_1=1$, $h_2=-0.00044$, $h_3= -22.08711$ and $\alpha=-0.00402$.
Importantly, the corresponding 2nd order of analytical approximate solutions are determined as
\begin{eqnarray}
A_z(z)=&&0.4572667 + 0.01498763 z + z^2 - 1.451489 z^3 - 0.005843966 z^4 - 0.06132506 z^5 \nonumber\\
&&+ 0.02351596 z^6 - 0.3336103 z^7 + 0.5323985 z^8 - 0.6967972 z^9 + 1.750668 z^{10} \nonumber\\
&&- 1.817722 z^{11} + 1.820144 z^{12} - 2.660147 z^{13} + 1.822905 z^{14} - 197263 z^{15} \nonumber\\
&&+ 1.916911 z^{16} - 2.772337 z^{17} + 4.148017 z^{18} - 5.315132 z^{19} + 6.836533 z^{20} \nonumber\\
&&- 8.412892 z^{21} + 8.246583 z^{22} - 8.138851 z^{23} + 7.799650 z^{24} - 5.610461 z^{25} \nonumber\\
&&+ 4.227755 z^{26} - 3.541047 z^{27} + 1.900061 z^{28} - 0.9300943 z^{29} + 0.6721495 z^{30} \nonumber\\
&&- 0.2165210 z^{31} - 0.007895318 z^{32},
\end{eqnarray}
\begin{eqnarray}
B_z(z)=&&0.4572613 + 0.001838552 z + 1.000001 z^2 - 1.453246 z^3 - 0.005859518 z^4 \nonumber\\
&&-2.118835\times 10^{-5}z^8+4.558038\times 10^{-5}z^9-7.453177\times 10^{-5}z^{10}\nonumber\\
&&+1.929635\times 10^{-4}z^{11}-2.202587\times 10^{-4} z^{12}+3.020311\times 10^{-4}z^{13}\nonumber\\
&&-3.686329\times 10^{-4}z^{14}+2.159995\times 10^{-4}  z^{15}-3.031449\times 10^{-5}z^{16}\nonumber\\
&&-2.347457\times 10^{-4}z^{17}+4.985287\times 10^{-4}z^{18}-7.088783\times 10^{-4}  z^{19}\nonumber\\
&&+6.819188\times 10^{-4} z^{20}-2.785920\times 10^{-4}z^{21} + 5.640031\times 10^{-5}z^{22} \nonumber\\
&&-9.993102\times 10^{-5}  z^{23}+ 7.812429\times 10^{-5}z^{24}-7.631288\times 10^{-5}z^{25}\nonumber\\
&&+1.244101\times 10^{-4}z^{26}-1.403398\times 10^{-4}  z^{27} + 1.307204\times 10^{-4}z^{28}\nonumber\\
&&-9.834623\times 10^{-5}z^{29} + 4.387319\times 10^{-5}z^{30} -1.459050\times 10^{-5}z^{31}\nonumber\\
&&+1.058479\times 10^{-5}z^{32},
\end{eqnarray}
\begin{eqnarray}
  \phi_z(z)=&&0.03217793 z + 0.07647953 z^3 + 0.0001430922 z^4 - 0.0003676241 z^5\nonumber\\
   &&- 0.0002630361 z^6- 0.00002948385 z^7 - 0.004393940 z^8 + 0.002189734 z^9 \nonumber\\
   &&- 0.008413222 z^{10}+ 0.007363797 z^{11} - 0.003525100 z^{12} + 0.002154606 z^{13} \nonumber\\
   &&+ 0.01131688 z^{14}- 0.01319623 z^{15} + 0.01786729 z^{16} - 0.009802537 z^{17} \nonumber\\
   &&- 0.005691524 z^{18}+ 0.004912230 z^{19} - 0.004087058 z^{20} + 0.0006557677 z^{21} \nonumber\\
   &&+ 0.004245723 z^{22}- 0.008432530 z^{23} + 0.01134451 z^{24} - 0.008496555 z^{25} \nonumber\\
   &&+ 0.003261786 z^{26}+ 0.004689052 z^{27} - 0.01143095 z^{28} + 0.01200959 z^{29} \nonumber\\
   &&- 0.008546128 z^{30}+ 0.002794749 z^{31}+6.702130\times 10^{-5} z^{32}.
\end{eqnarray}
Now, we can compare the analytic approximate solutions with the numerical solutions appeared in the previous section.
We plot the analytic approximate solutions ($A_z^{\rm ana}$, $B_z^{\rm ana}$ and $\phi_z^{\rm ana}$) and
numerical solutions ($A_z^{\rm num}$, $B_z^{\rm num}$ and $\phi_z^{\rm num}$) in Figure~\ref{fig33}
for $r_h=1$, $\lambda=0.892$, and $-\Lambda/3=0.457$. They are apparently consistent with each other.

\vspace{-12pt}

\begin{figure*}[t!]
\centering
\includegraphics{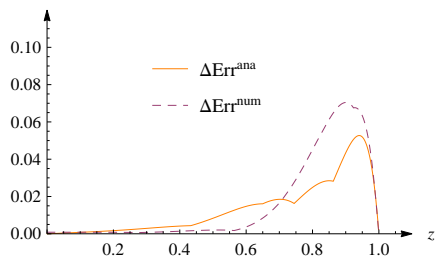}
\caption{ Total absolute errors for analytic approximate solutions $(\Delta \text{Err}^{\rm ana})$ and numerical
solutions $(\Delta \text{Err}^{\rm num})$ from three field equations.}\label{fig4}
\end{figure*}

It is interesting to check the accuracies of these analytic approximate solutions
and numerical solutions based on the field Equations \eqref{eq1z}--\eqref{eq3z}. Substituting these solutions
into Equations~\eqref{eq1z}--\eqref{eq3z}, the total absolute errors from three field equations are obtained as
\begin{eqnarray}
\Delta \text{Err}\equiv \left|\Delta eq_1\right|+\left|\Delta eq_2\right|+\left|\Delta eq_3\right|. \label{error}
\end{eqnarray}
Total absolute errors of these analytic approximate and numerical solutions are displayed in Figure \ref{fig4}.
One can find that the total absolute error for analytic approximate solutions is smaller than that for the numerical solutions.
In other words, the analytic approximate solutions are more accurate than the
numerical solutions when solving nonlinear Equations \eqref{eq1z}--\eqref{eq3z}.

\begin{figure*}[t!]
\centering
\includegraphics{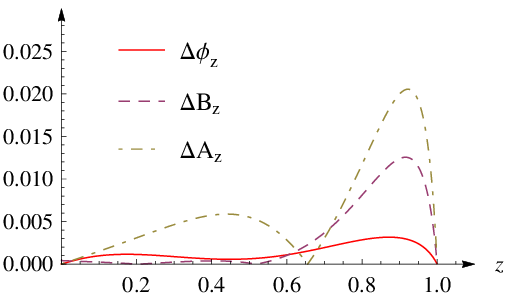}
\hfill%
\includegraphics{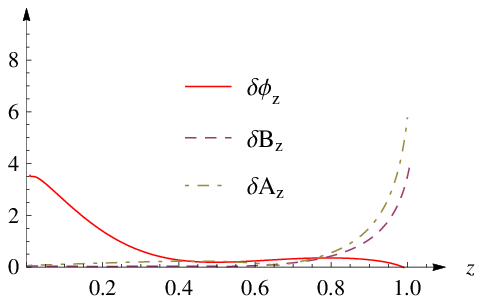}
\caption{Absolute (Left) and relative (Right) errors between numerical and analytical approximate solutions.}\label{fig5}
\end{figure*}

To comparing these solutions, we further calculate the absolute differences between numerical
and analytical approximate solutions by taking
\begin{eqnarray}
\Delta A_z=\left|A_z^{num}-A_z^{ana}\right|,\quad \Delta B_z=\left|B_z^{num}-B_z^{ana}\right|,\quad
\Delta\phi_z=\left|\phi_z^{num}-\phi_z^{ana}\right|.
\end{eqnarray}
The relative errors can be also calculated in the form of
\begin{eqnarray}
&&\delta A_z=\frac{\left|A_z^{num}-A_z^{ana}\right|}{A_z^{num}}\times100\%,\quad \nonumber\\
&&\delta B_z=\frac{\left|B_z^{num}-B_z^{ana}\right|}{B_z^{num}}\times100\%,\quad \nonumber\\
&&\delta\phi_z=\frac{\left|\phi_z^{num}-\phi_z^{ana}\right|}{\phi_z^{num}}\times100\%.
\end{eqnarray}
We find that main differences between numerical solutions and analytic approximation solutions occur close to
the event horizon for metric functions $A(z)$ and $B(z)$ and region far from the black hole for scalar
field function $\phi(z)$, see Figure \ref{fig5}.

\vspace{-12pt}

\section{Conclusions and discussions} \label{sec5}

In this work, we investigated the spontaneous scalarization for SAdS black holes  thoroughly in ESGB theory.
The SAdS black holes become prone to tachyonic instability triggered by the strong space-time curvature in some
region of the parameter space. Then, scalarized AdS black holes could emerge from SAdS black holes at bifurcating points.
Numerical solutions for scalarized AdS black holes are obtained for $\lambda=0.892$ and \mbox{$-\Lambda/3=0.457$}.

Later, we derive the analytical approximate solutions for metric functions $A(z)$ and $B(z)$ and scalar field $\phi(z)$ by using the HAM. The region and rate of convergence
of the series solution for the HAM does not depend on the choice of the initial guess function,
auxiliary linear operator, and an auxiliary function, but it can be effectively controlled by using a convergence control parameter.
Since the approximation is significantly accurate in the entire space-time outside the event horizon, it can be used for studying
the properties of this particular black hole and the various phenomena.  The present work is considered as an important work
because we confirm that numerical solutions are consistent with an analytical approximate solution for the scalarized AdS  black hole.

As an avenue of a further research, one may propose the related properties of scalarized AdS black hole (thermodynamics,
Hawking radiation, particle motion, shadow, stability and QNMs) and compare them with SAdS black holes.

 \vspace{1cm}

{\bf Acknowledgments}
 \vspace{1cm}

We appreciate Rui-Hong Yue for helpful discussion. D. C. Z acknowledges financial support
from Outstanding Young Teacher Programme from Yangzhou University, No. 137050368. M. Y. L acknowledges financial
support from the Initial Research Foundation of Jiangxi Normal University.

\end{document}